\documentclass[doublecol]{epl2} 
\usepackage{graphicx}
\usepackage{epsfig}
\usepackage{amssymb}
\usepackage{amsmath}
\usepackage{bm}
\usepackage{epstopdf}

\begin{document}

\title{Fractional energy states of strongly-interacting bosons in one dimension}
\shorttitle{Fractional energy states of strongly-interacting bosons in one dimension}

\author{N.~T. Zinner\inst{1} \and A.~G. Volosniev\inst{1} \and D.~V. Fedorov\inst{1} \and A.~S. Jensen\inst{1} \and M. Valiente\inst{2}}

\institute{                    
  \inst{1} Department of Physics and Astronomy, Aarhus University, DK-8000 Aarhus C, Denmark\\
  \inst{2} SUPA, Institute for Photonics and Quantum Sciences, Heriot-Watt University, Edinburgh EH14 4AS, United Kingdon}

\date{\today}

\abstract{
We study two-component bosonic systems with strong 
inter-species and vanishing intra-species interactions. 
A new class of exact eigenstates is found with energies 
that are {\it not} sums of the single-particle energies 
with wave functions that have the characteristic 
feature that they vanish over extended regions of coordinate space.
This is demonstrated in an analytically solvable model for three
equal mass particles, two of which are identical bosons, which is exact in the strongly-interacting
limit. We numerically verify our results by presenting the first application of the stochastic variational 
method to this kind of system. We also demonstrate that the limit where both inter-
and intra-component interactions become strong must be treated with extreme care as these limits do not
commute.
Moreover, we argue that such states are generic also for general multi-component systems with 
more than three particles.
The states can be probed using the same techniques that have recently been used
for fermionic few-body systems in quasi-1D.}

\maketitle

\section{Introduction}
Ultracold atomic gas experiments have proven an invaluable tool for 
realizing strongly-correlated quantum mechanical systems in highly 
tunable environments \cite{bloch2008}. A prominent example is 
the so-called Tonks-Girardeau (TG) gas \cite{tonks1936,girardeau1960} 
of impenetrable bosons in 
one dimension (1D) that has been created using cold
atoms \cite{paredes2004,kinoshita2004,kinoshita2005,haller2009}. An exciting
recent advance in this direction is the ability to produce and 
manipulate low-dimensional samples with controllable particle
numbers down to single digits \cite{he2010,serwane2011,zurn2012,bourgain2013,wenz2013}.
These developments show that few-body systems with bosons and fermions in 
microtraps that can be manipulated and studied in great detail can 
be achieved with ultracold atoms. These systems would facilitate
access to strongly correlated states with applications in 
quantum information, computation, and atomtronics 
\cite{anderlini2007,folling2007,trotzky2008,negretti2010,simon2011,bakr2011,schneider2012}.

One-dimensional quantum systems
have served as playgrounds for many theorists due to the
presence of exact solutions.  Most of these are built on the 
Bethe ansatz first introduced for studying magnetism in 1D metals \cite{bethe1931}. The new
possibilities for trapping cold atoms with tunable 
short-range interactions in effective 1D geometries has 
generated frenetic recent activity
\cite{girar2004,girardeau2007,zollner2008,tempfli2009,deuret2008,girardeau2011,valiente2012,garcia2013}.
While many of these works use different generalizations of the 
original Bose-Fermi mapping of Girardeau \cite{girardeau1960}, it was
recently shown that the mapping fails for the case of trapped 
two-component Fermi gases already at the few-body level 
\cite{chen2009,lindgren2013,sowinski2013,cui2013,gharashi2013}, ushering in the need
for a more general technique to address multi-component fermionic 
systems \cite{volosniev2013}. This begs the question of whether
there could be some overlooked features of two-component Bose
systems in the strongly-interacting regime.
A recent numerical study \cite{garcia2013}, suggests that there is 
a non-trivial crossover between the composite fermionized 
regime (weak intra-component and strong inter-component interactions 
\cite{zollner2008}) and a regime of phase separation when one
of the components attains strong inter-component repulsion.

In this paper we describe a class of states for two-component bosons
with strong short-range interactions that is different
from those obtained by a
Bose-Fermi mapping to spinless fermions \cite{girardeau1960}.
We demonstrate this using a model for three 
harmonically trapped equal 
mass particles where two are identical bosons of type
$A$ and the third of a distinct type, $B$.  When the 
$A$ bosons are non-interacting, 
we find two types of eigenstates
when the $AB$ interaction becomes strong; one
set can be related to the wave function of spinless
fermions whereas the others (including the ground state)
have wave functions that 
are highly correlated and cannot be obtained or built by
a mapping to fermions. This could be regarded as natural
when one species is non-interacting. However, we show that even in this
case a subset of the spectrum can in fact be obtained by a Bose-Fermi 
mapping.
In sharp constrast to 
states related to spinless fermions,
the new class of states will generally have
energy eigenvalues that are not integer multiples of
the harmonic oscillator energy unit (disregarding 
zero-point energies). 
We confirm the 
analytical finding by a stochastic variational 
calculation which is, to the best of our knowledge,
the first time this technique has been applied to 
strongly-interacting one-dimensional systems.
Furthermore, we show that the limit where both 
$AA$ and $AB$ interactions become strong is 
very delicate and yields different eigenstates 
depending on the order in which the couplings are
taken to infinity.
As we demonstrate below, 
our findings imply that general 
multi-component $N$-boson systems will have such 
solutions and they must be considered when 
addressing strongly-interacting 1D bosonic systems.
We also show that current experimental techniques using
either tunneling out of a trap or RF spectroscopy should
be able to see a clear distinction between the 
integer and fractional energy states in these
strongly-interacting systems. 

\begin{figure}[t!]
\centering
\includegraphics[scale=0.42]{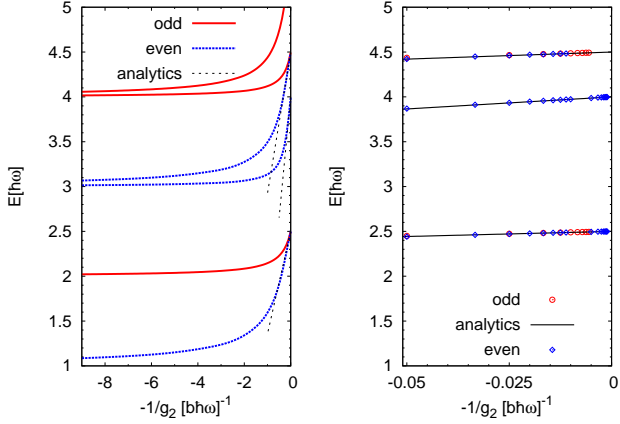}
\caption{Energy spectra obtained using the stochastic variational method. Left panel shows
the energy on for repulsive $g_2>0$ interactions for the lowest even and odd states in the spectrum.
The right panel shows a zoom of the same data around $g_{2}^{-1}=0$ and includes analytical results for 
the energy to first order in $g_{2}^{-1}$. We clearly see the analytical and numerical results merge
close to resonance, proving the convergence of our calculations.}
\label{energies}
\end{figure}

\section{Model}
We consider two identical bosons ($A$) with coordinates $x_1$ and $x_2$
and a third particle ($B$) with coordinate $x_3$ which is distinct from the 
bosons but of the same mass, $m$. This can be realized using bosons with two different
internal (hyperfine) states in the context of cold atoms \cite{bloch2008}. The particles move
in a harmonic oscillator with frequency 
$\omega$ and oscillator length $b=\sqrt{\hbar/m\omega}$ and have
short-range pair
potentials that we model as delta functions, i.e.
\begin{equation}
V=g_1\delta(x_1-x_2)+g_2\delta(x_1-x_3)+g_2\delta(x_2-x_3).
\end{equation}
Defining the coordinates $x=(x_1-x_2)/\sqrt{2}$, $y=(x_1+x_2)/\sqrt{6}-\sqrt{2/3}x_3$, 
and $R=(x_1+x_2+x_3)/\sqrt{3}$, we may separate the center-of-mass, $R$, and consider
the relative wave function, $\Psi(x,y)$. Bosonic symmetry requires $\Psi(-x,y)=\Psi(x,y)$
and since our system is parity invariant, the parity is thus determined by
the sign of $\Psi(x,y)$ when $y\to -y$. This means that once we 
have the solution for $x>0$ and $y>0$, the full solution can be obtained by 
continuation using parity and Bose symmetry.
Away from the points at which two particles meet, the solutions must be eigenfunctions
of the free Hamiltonian for three particles in a harmonic trap. The two 
regular normalizable solutions are \cite{harshman2012}
\begin{equation}\label{generalwave}
\Psi(\rho,\phi)=N \rho^\mu U(-\nu,\mu+1,\rho^2)e^{-\rho^2/2}
\left\{\begin{matrix}\cos(\mu \phi)\\ \sin(\mu\phi)\end{matrix}\right.,
\end{equation}
where $U(a,b,x)$ is the Tricomi function, $\rho=\sqrt{x^2+y^2}$ is the hyperradius, 
$\phi=\textrm{arctan}(y/x)$ the hyperangle, and $N$ is a normalization 
constant. 
The corresponding energy
is $E=\hbar\omega(2\nu+\mu+1)$. 

The interactions can be implemented by matching solutions in different regions
of space through continuity of the wave function and the conditions
\begin{equation}
\frac{\hbar^2}{2m\rho^2}\left(\frac{d\Psi(\rho,\phi)}{d\phi}|_{\phi_0+\epsilon}-\frac{d\Psi(\rho,\phi)}{d\phi}|_{\phi_0-\epsilon}\right)
=\frac{g_i}{\sqrt{2}\rho}\Psi(\rho,\phi_0),
\label{boundary}
\end{equation}
for any $\rho$ and where $\phi_0$ is an angle where two particles
overlap ($i=1$ for $AA$ overlap and $i=2$ for $AB$ overlap).
We now define
rescaled coupling strengths, $\tilde g_i=\sqrt{2}m\rho g_i/\hbar^2$. In terms of the 
$\tilde g_i$, Eq.~\eqref{boundary} is now independent of $\rho$ and we have achieved 
an effective decoupling of the radial and angular equations. 
The decoupling means that we get an equation that is independent of $\nu$.
The crucial point is that
when either $g_1\to\infty$ and $g_2=0$, $g_2\to\infty$ and $g_1=0$, or $g_1=g_2\to\infty$, 
our model is exact. Our model can thus be used to 
obtain the exact wave functions and energies in limits where one or both couplings
are large.

\begin{figure}[t!]
\centering
\includegraphics[scale=0.42]{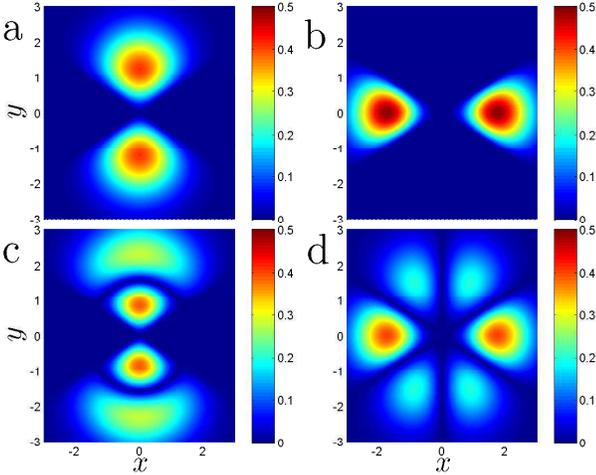}
\caption{Contour plots of the (absolute) value of the wave function in the 
plane of relative coordinates $x$ (horizontal) and $y$ (vertical). The 
systems have two identical non-interacting particles that both interact
with a third particle of the same mass with a zero-range interaction of 
infinite strength. The upper left
panel is the even parity ground state and the upper right is the even parity
first excited state, while the lower left panel is the second excited state also
for even parity. For comparison, the lower right panel shows the case of two 
identical fermions and a third particle (of the same mass).}
\label{con2D}
\end{figure}

The even parity solutions can be obtained by assuming that the angular wave function, $F_i(\phi)$,
has the form, $F_1(\phi)=\alpha\cos(\mu\phi)$, an even function in $\phi$ and thus in $y\to -y$, 
for $0<\phi<\pi/6$. In the region $\pi/6<\phi<\pi/2$, we assume the general form 
$F_2(\phi)=\beta\sin(\mu\phi+\delta)$. $\alpha$, $\beta$, and $\delta$ are constants.
Using continuity
and Eq.~\eqref{boundary} at $\phi_0=\pi/6$ and $\phi_0=\pi/2$ gives the 
eigenvalue equation that determines $\mu$,
\begin{equation}
-\mu\cot\left(\frac{\pi}{3}\mu+\textrm{arctan}(\frac{2\mu}{\tilde g_1})\right)
+\mu\tan\left(\frac{\pi}{6}\mu\right)-\tilde g_2=0.
\label{ene-even}
\end{equation}
When $g_1=0$ and $g_{2}\to \infty$ our model is exact and 
we have thus obtained a whole class
of solutions indexed by $\nu$. 
The solutions with odd parity can be obtained in similar 
fashion by exchanging sine and cosine in $F_i(\phi)$ to obtain
\begin{equation}
\mu\tan\left(\frac{\pi}{3}\mu-\textrm{arctan}(\frac{\tilde g_1}{2\mu})\right)
-\mu\cot\left(\frac{\pi}{6}\mu\right)-\tilde g_2=0.
\label{ene-odd}
\end{equation}
In the limit $\tilde g_{2}^{-1}=0$ and $\tilde g_1=0$, Eqs.~\eqref{ene-even} and \eqref{ene-odd}
have two types of solutions; integer for $\mu=3+6n$ with even and $\mu=6n$ with odd parity, 
and surprisingly also half-integer for
$\mu=3/2+3n$ for odd and even parity. Here $n>0$ is an integer. 
If we subsequently let
$\tilde g_1\to \infty$, then only integer solutions remain. The 
analytical model predicts that this process is smooth, i.e. the 
half-integer solutions go continuously to integer $\mu$. This implies
that non-integer solutions are a generic feature of multi-component
bosons. This yields analytical insight to the findings of Ref.~\cite{garcia2013}.
In the absence of a trapping potential, the Bethe ansatz
equations for multi-component systems has been discussed
\cite{sutherland1968}. 
The three-body bound state for $g_1=0$ 
was discussed by Gaudin and Derrida \cite{gaudin1975}. However, 
the general case in the limit $g_1\ll g_2$ is very rarely discussed 
in the literature.

\section{Numerical solutions}
In order to verify the analytics, we 
use stochastic variational calculations \cite{suzuki1998,svm2013} with 
$g_1=0$ for repulsive $g_2>0$. The spectrum is shown in Fig.~\ref{energies} and
confirms the analytical spectrum
for $g_{2}^{-1}\to 0^+$. Our work therefore demonstrates 
the applicability of the stochastic variational method to strongly-interacting
problems in 1D. The analytical wave functions
at $g_{2}^{-1}=0$ can be used to calculate the energy to linear order in $g_{2}^{-1}$, 
i.e. $E=E_0-\tfrac{K}{g_2}$ (see the appendix for details). 
The slope, $K$, can then be used  
to show convergence as done 
in Fig.~\ref{energies}. Note the equal slopes at $g_{2}^{-1}=0$ for
fractional energy states of both parities.
At integer energies, 
the odd and even solutions never become degenerate (irrespective of $g_1$)
as can be easily checked from Eqs.~\ref{ene-even} and \ref{ene-odd}.

\begin{figure}[t!]
\centering
\includegraphics[scale=0.42]{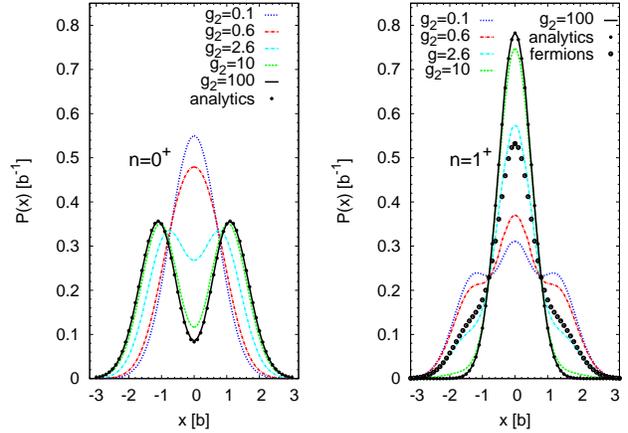}
\caption{Density of the single $B$ particle for $g_1=0$ for different values of the interparticle
interactions in the ground state (left) and first excited state (right) for even parity. Similar
results can be obtained for odd parity states. On the right panel we also plot the results
for a system where the two identical particles are fermions in the infinite interaction strength
limit.}
\label{density}
\end{figure}

\section{Fractional energy states}
To gain further insight, we show contour plots of the 
wave functions at $g_{2}^{-1}=0$ as functions of the relative
coodinates $x$ and $y$. Panels
a, b, and c in Fig.~\ref{con2D} display even parity 
ground ($E=2.5\hbar\omega$), first ($E=4\hbar\omega$), 
and second excited ($E=4.5\hbar\omega$) states, respectively.
The second excited state has $\mu=3/2$ and $\nu=1$, i.e. 
it contains a radial excitation (visible in Fig.~\ref{con2D} 
panel c 
by the node along the vertical axis). 
Most striking 
is the presence of regions where the wave functions 
are zero, and moreover that these regions are complementary for 
the fractional and integer solutions. This immediately 
implies that these states are {\it very} different in spatial
structure.
The fractional states only allow the three particles
to be in configurations where the $B$ coordinate, $x_3$, is either 
larger or smaller
than both $x_1$ and $x_2$ (so that $B$ is found either on 
the left or right of {\it both} $A$ particles). The corresponding
density is shown on the left in Fig.~\ref{density}. For the integer
states particle $B$ will always be located between the two $A$ particles, 
i.e. $x_1<x_3<x_2$ or $x_2<x_3<x_1$ as shown on the right in 
Fig.~\ref{density}. For comparison, we show in 
Fig.~\ref{con2D} panel d the wave function when the 
$A$ particles are identical fermions instead \cite{lindgren2013,gharashi2013}
and the corresponding density on the right in Fig.~\ref{density}.
The integer energy states with $g_1=0$ and $g_{2}^{-1}=0$ 
can be built from spinless fermion wave functions.

The reason for the half-integer energy can be seen from 
Fig.~\ref{con2D} panel a. The wave function must be symmetric 
in $x$ by Bose symmetry and vanish when $x_1=x_3$ and $x_2=x_3$.
The 
angular part in Eq.~\eqref{generalwave} must therefore advance its argument 
by half a period in the 120 degree wedge given by $\pi/6<\phi<5\pi/6$ (or equivalently
$7\pi/6<\phi<11\pi/6$), which means that we have
$(2n-1)\pi=2\pi\mu/3$ or $\mu=3(2n-1)/2$ with integer $n>0$. 
This can be related to classical scattering from wedges in free 
space \cite{ober1958}. 
Here a 120 degree wedge (corresponding to our fractional states) 
is diffractive while a 60 degree wedge (corresponding to our integer states) 
is not.
The diffractionless case implies the existance of a
free fermion model and that a Bethe ansatz solution 
should only be expected in the latter case. Indeed it has been shown 
recently \cite{lama2013} that $g_1\neq g_2$ leads to diffractive 
scattering. We can thus understand our findings as a signature
of diffractive and non-diffractive states. In quantum mechanical 
terms, the distinction can be seen clearly in the three-body 
wave functions. Due to the factor $\rho^\mu$, it 
can {\it only} be expanded in terms of a finite number of
single-particle wave functions when $\mu$ is integer (and 
thus mapped to a free fermion model). For non-integer
$\mu$ this is impossible.
This makes the fact that we have
a closed formula for a class of the latter type even more interesting.

\section{Order of limits}
\begin{figure}[t!]
\centering
\includegraphics[scale=0.42]{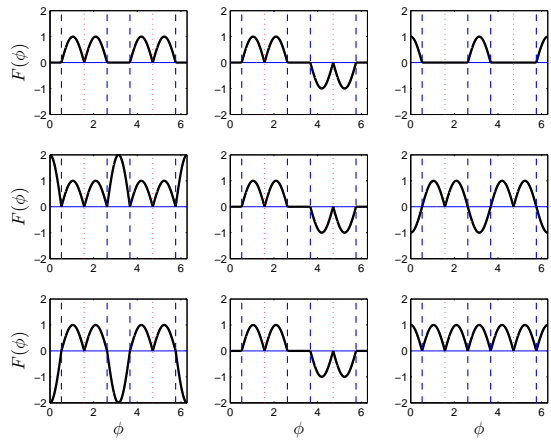}
\caption{Angular wave functions, $F(\phi)$, for the entire angular range, $0\leq \phi\leq 2\pi$, of 
the exact solutions at $E=4\hbar\omega$ obtained by taking the limits of diverging $g_1$ and $g_2$ 
in different ways. The three top panels shows the case where $g_{2}^{-1}=0$ and then we take $g_1\to\infty$,
the three middle panels are obtained when $g_{1}^{-1}=0$ and then $g_{2}\to\infty$, while the three
lower panels are obtained for $g_1=g_2\to\infty$. Vertical dashed (blue) lines are at the overlaps 
of $A$ and $B$, while dotted (red) lines are at the overlaps of $A$ and $A$ where Bose 
symmetry applies. Left and right
columns are even parity states, while the middle column has odd parity. Note the orthogonality
of the states in each row.
}
\label{limits}
\end{figure}

An interesting question with a surprising answer concerns the limit
where both $g_1$ and $g_2$ become large. This can be addressed by using
the method introduced in Ref.~\cite{volosniev2013}. The full analytical details can be 
found in the appendix below. The wave function must now 
vanish when any two particles overlap, $\mu$ goes to an integer
($\mu=3$ for the lowest state).
A physically motivated way to 
see that the order of limits 
matters is to consider states like
those in Fig.~\ref{con2D} panels a and b which have $g_1=0$ and
$g_{2}^{-1}=0$. Increasing $g_1$, the state in panel b
is unaffected (no amplitude on the $y$ axis) while the ground
state in panel a feels the interaction and goes to $E=4\hbar\omega$
as $g_1\to\infty$.
The upper row in Fig.~\ref{limits} shows
the three resulting angular wave functions. In contrast, starting
from $g_{1}^{-1}=0$ and $g_2=0$, the fractional 
energy state is absent since a node is required along the $y$-axis. 
Taking now the limit $g_2\to\infty$ produces the wave function in 
the middle row of Fig.~\ref{limits}. All the cases in the middle
row
have absolute value of the wave functions and 
densities identical to the case where the $A$ particles
are fermions (shown in Fig.~\ref{con2D}d and on the right in Fig.~\ref{density})
and are thus directly related to the three-body solution for 
two-component fermions discussed in Ref.~\cite{guan2009}.
Finally, when $g_1=g_2\to\infty$ we obtain the bottom row in Fig.~\ref{limits}.
The far right state which is obtained by using the original 
Bose-Fermi mapping of Girardeau \cite{girardeau1960}, but two other
states emerge. This shows the importance the order
of limits in analytical and numerical calculations. 
Note that all the states in Fig. 4 have energy
$E=4\hbar\omega$, and we thus recover the three-fold 
degeneracy in the strongly interacting limit 
\cite{deuret2008,girardeau2007,volosniev2013}.

\section{Detection}
The states can be detected by
tunneling experiments similar to those used for fermions \cite{zurn2012,zurn2013}.
For $g_1=0$ and $g_2$ large, we may assume that $A$ and $B$ atoms cannot 
penetrate each other. Thus only the atom located on the 
side of the trap where it is opened can tunnel. The fractional 
state has equal probabilities of spatial $AAB$ and $BAA$ ordering, so
50\% of tunnel experiments will produce an $AA$ final state and 50\% 
will produce $BA$. For integer states, we have $ABA$ structure and 
always produce an $AB$ final state. 
Fig.~\ref{density} show the two situations and demonstrates that 
$g_2\geq 10$ is already close to the fermionized limit.
One can also use RF spectroscopy \cite{wenz2013} to probe
the states by starting from a weakly-interacting $AAA$
system and applying a pulse to drive from internal state $A$ to $B$
in a regime of strong $AB$ interaction. When the driving frequency
matches a fractional energy, this should produce the state with 
an equal probability of $AAB$ and $BAA$ configurations, while
if it matches the integer energy we get a pure $ABA$ wave function. 
The same holds if two atoms are converted from $A$ to $B$ due to the 
symmetry of the problem. With well-studied
Feshbach resonances and previous realizations of a spinor gas 
we suggest $^{87}$Rb for experimental realization \cite{kurn2013}.

\section{Larger systems}
To show that fractional states appear also for larger systems, we
consider three 
non-interacting $A$ particles and one $B$. 
Assume that $g_{2}^{-1}=0$
and assume a wave function built from spinless fermions.
Since any ordering will always have two adjacent $A$ particles, 
Bose symmetry implies that we get a cusp when two $A$ particles 
coincide
(absolute value of a spinless fermion wave function).
This is not allowed when $g_1=0$ and we cannot use spinless
fermion wave functions. The argument generalizes to even
larger systems. However, structures like f.x. $ABAB$ and $ABABA$
are still allowed and generally we will have both integer and
fraction classes of exact eigenstates. This also applies to Bose-Fermi mixtures. 
As a means for detecting fractional and integer 
states for larger systems, we envision a measurement of the 
momentum distribution which is sensitive to 
nodes and cusps in the wave function \cite{deuret2008}. Since 
the nodal structure of states like $AAABBB$ and $ABABAB$ are
very different we expect distinctly different momentum distributions.

\section{Outlook}
The fractional solutions are unique correlated states with 
an interesting diffractive classical counterpart. They can provide
an excellent benchmark for various numerical procedures
and demonstrate the care with which strong interacting limits must 
be handled. Combined with recent results on 1D 
fermionic systems \cite{lindgren2013,volosniev2013}, we now
have a full analytical classification of strongly-interacting two-component three-body systems. 

This work was supported by the Sapere Aude program starting grant 
under the Danish Council for Independent Research and by a project grant from 
the Danish Council for Independent Research - Natural Sciences.

\section{Appendix - Slope of the energy at fermionization}
To obtain the slope of the energy in the different strongly interacting limits, we follow the 
approach outlined in Ref.~\cite{volosniev2013} and use a linear expansion in $1/g$, where
$g$ is the strength of a delta function two-body interaction and we will assume that $g>0$ here. 
Writing $E=E_0-K/g$, we have $K=-\left[\frac{\partial E}{\partial g^{-1}}\right]_{g^{-1}=0}$.
\begin{align}
K=\lim_{g\to\infty}
g^2\frac{\sum_{i<j}\int\prod_{i=1}^{3}dx_i\left|\Psi(x_1,x_2,x_3)\right|^2
\delta\left(x_i-x_j\right)}{\langle \Psi|\Psi\rangle},
\label{slope}
\end{align}
By using the appropriate boundary condition for a delta-function potential, 
one can show that $g$ drops out of the calculation of $K$ which depends 
only on $\Psi$ \cite{volosniev2013}.

First we show how to obtain the slopes shown in Fig.~1 of the main text from the wave functions that we obtain
in the analytical model. This is a straighforward matter of inserting those wave functions into the equation for
$K$ found in Ref.~\cite{volosniev2013}.
Ignoring the center-of-mass part, the wave functions can be written $\Psi(\rho,\phi)=P(\rho)F(\phi)$. 
The radial parts for the three lowest even parity states shown in Fig.~1 of the main text at $g_{1}^{-1}=0$ are
$P_0(\rho)=\rho^{3/2}L_{0}^{3/2}(\rho^2)e^{-\rho^2/2}$, 
$P_1(\rho)=\rho^{3}L_{0}^{3}(\rho^2)e^{-\rho^2/2}$ and
$P_2(\rho)=\rho^{3/2}L_{1}^{3/2}(\rho^2)e^{-\rho^2/2}$
where $L_{n}^{m}(x)$ is the associated Laguerre polynomial which arises
since $U(-n,m+1,x)=(-1)^n n! L_{n}^{m}(x)$ for integer $n$ in Eq.~(2) of the main text.
The angular functions for $\mu=3/2$
are $\sin(3/2(\phi-\pi/6))$ for $\pi/6<\phi<5\pi/6$, 
$\pm\sin(3/2(\phi-7\pi/6))$ for $7\pi/6<\phi<11\pi/6$ and zero otherwise. The
$\pm$ sign is what separates the opposite parity states that become degenerate
at $\mu=3/2$.
The 
excited state with $\mu=3$ has angular functions $\cos(3\phi)$ for 
$0<\phi<\pi/6$ or $11\pi/6<\phi<2\pi$, $-\cos(3\phi)$ for 
$5\pi/6<\phi<7\pi/6$ and zero otherwise. Using
these wave functions, we obtain
$K_0=\frac{9}{\sqrt{2\pi^3}}$,
$K_1=\frac{27}{4\sqrt{2\pi}}$ and
$K_2=\frac{117}{20\pi^{3/2}}$.
These are the values used to generate the linear fits in Fig.~\ref{energies}. 
This is 
not a surprise as the wave functions in the two cases are very different as shown in panel b and 
panel d of Fig.~\ref{con2D}.

\section{Appendix - Limit of strong intra- and inter-component repulsion}
We now address the different ways in which to approach the limit where both $g_1$ and $g_2$ go to
infinity. This immediately implies that the three-body wave function must be zero whenever any
pair of particles overlap. This is only possible using the integer $\mu$ solutions. We
concentrate on the ground state(s) which have $\mu=3$. This means that we are in the 
case where we can use spinless wave functions to express $K$ as a functional of a set of 
coeffecients $a_i$ of the spinless wave function in each region of space of given ordering 
of the three coordinates $x_1$, $x_2$, and $x_3$. This technique is outlined in Ref.~\cite{volosniev2013}
for the case of fermions. The generalization to bosons is a straightforward matter of 
applying Bose symmetry. Just as in the case of two identical fermions, we may reduce this
set of parameters by using parity and Bose symmetry to three. The wave function is then 
$a_1\Psi_F$ for $x_2<x_1<x_3$ ($\pi/6<\phi<\pi/2$), $a_2\Psi_F$ for $x_2<x_3<x_1$ ($0<\phi<\pi/6$ and $11\pi/6<\phi<2\pi$), 
and $a_3\Psi_F$ for
$x_3<x_1<x_2$ ($3\pi/2<\phi<11\pi/6$). The other orderings are dictated by symmetries. 
Here the antisymmetrized product state is
$\Psi_F\propto \exp(-\frac{\rho^2}{2}-\frac{R^2}{2})\rho^3 \cos(3\phi)$. The
$F$ indicates the common name 'fermionized' state.

We start with the case where
we assume that $g_{2}^{-1}=0$ and then take the limit $g_{1}\to\infty$. 
Since $g_{2}^{-1}=0$, the interaction term with $g_2$ is 
excluded when calculating the slope $K$ from Eq.~\eqref{slope}. It 
implies the boundary condition that the wave function
must vanish when non-identical particles coincide. The expression 
for the slope when $g_1\to\infty$ is then determined by the 
derivatives from points where the two identical bosons overlap. It can 
be written
\begin{align}
K=2\gamma \frac{a_{1}^{2}}{a_{1}^{2}+a_{2}^{2}+a_{3}^{2}}=2\gamma\frac{a_{1}^{2}}{2a_{1}^{2}+a_{2}^{2}},
\label{g1slope}
\end{align}
where in the second equation we use parity and Bose symmetry which implies that $a_1=a_3$. 
The factor $\gamma$ is independent of the $a_i$ coefficients and depends only on the 
fermionized state $\Psi_F$. It has the expression 
\begin{align}
\gamma=\frac{\int_{x_1<x_2}\left|\left[\left(\frac{\partial}{\partial x_2}-\frac{\partial}{\partial x_3}\right)_{x_3-x_2\to 0}\right]\Psi_F\right|^2\textrm{d}x_1\textrm{d}x_2}
{\int_{x_1<x_2<x_3}\textrm{d}x_1\textrm{d}x_2\textrm{d}x_3|\Psi_F|^2}.
\end{align}
This factor may be computed using the wave function $\Psi_F$. Since it is a prefactor it
does not influence the relation between the $a_i$ coefficients and we do not calculate it explicitly. 
As outlined in Ref.~\cite{volosniev2013}, one can now obtain the eigenfunctions 
in the vicinity of $g_{1}^{-1}=0$ by finding the extreme points of Eq.~\eqref{g1slope}. 
By differentiating Eq.~\eqref{g1slope} with respect to $a_1$ and $a_2$ and equating it to 
zero, one immediately sees that the two solutions are $a_2=0$ and $a_1$ arbitrary or 
$a_1=0$ and $a_2$ arbitrary. The arbitrary value is then fixed by normalization. 
The even parity solution with $a_2=0$ is the left panel in the top row of Fig.~4 in the main text, 
while the one with $a_1=0$ is the right panel. The odd parity solution is $a_2=0$ and $a_1=-a_3$
and is shown in the top middle panel in Fig.~\ref{limits}. Note that the solution 
with $a_1=0$ has $K=0$. This is expected since this solution has energy $E=4\hbar\omega$ 
for $g_1=0$ and thus its energy does not change as we take $g_1\to\infty$. This 
is clear also from the wave function in panel b of Fig.~\ref{con2D} since this state
has zero amplitude in the region where the $g_1$ interaction is located (along the 
$y$-axis).

The second case of interest is the one where we first take $g_{1}^{-1}=0$ and then 
let $g_2\to\infty$. Here we exclude the $g_1$ interaction term when calculating $K$ 
from Eq.~\eqref{slope} and
keep the ones with $g_2$. This produces the equation
$K=\gamma \tfrac{(a_1-a_2)^2}{2a_{1}^{2}+a_{2}^{2}},$
where again we have $a_1=a_3$ as before for positive parity. The extreme points of 
this equation are $a_1/a_2=-1/2$ (left panel in the middle row in Fig.~\ref{limits})
and $a_1/a_2=1$ (right panel in the middle row in Fig.~\ref{limits}). 

Finally, we consider the symmetric case where $g_1=g_2\to\infty$.
We retain all three interaction terms when calculating $K$, and from 
the point of view of the Hamiltonian all three particles are identical. This 
gives the expression
$K=\gamma \tfrac{(a_1-a_2)^2+2a_{1}^{2}}{2a_{1}^{2}+a_{2}^{2}},$
where again we have $a_1=a_3$ as before for positive parity. The extreme points of 
this equation are $a_1/a_2=1/2$ (left panel in the middle row in Fig.~\ref{limits})
and $a_1/a_2=-1$ (right panel in the middle row in Fig.~\ref{limits} in the main text). The 
change of sign between $a_1$ and $a_2$ for the completely symmetric solution 
occurs because $\Psi_F$ changes sign and this is compensated by $a_1=-a_2$
to yield the solution that is obtained by Girardeau's Bose-Fermi mapping of
the problem \cite{girardeau1960}. The odd parity state has $a_2=0$ 
(by Bose symmetry) and $a_1=-a_3$ (combined Bose and parity symmetry).
The odd state turns out to be the same irrespective of how the limit
is taken.

One can also use the slope to determined the adiabatic connections 
between strongly interacting and non-interacting states \cite{volosniev2013}. 
This requires calculation of $\gamma$ and 
insertion of the solutions for $a_i$ in $K$ for the different cases. 
In the symmetric case where $g_1=g_2\to\infty$, the non-interacting 
ground state is connected to the Girardeau state in the bottom right
panel of Fig.~\ref{limits}. In the two other cases the non-interacting
ground state (of even parity) is connected to the wave functions in 
the left panel of middle and top rows in Fig.~\ref{limits}. This is
a clear demonstration of the delicate nature of multi-component 
bosonic 1D systems with strong short-range interactions.

\end{document}